\documentclass[%
reprint,
amsmath,amssymb,
aps,prb,
twocolumn,
floatfix,
showpacs]
{revtex4-1}

\usepackage{graphicx}
\usepackage{dcolumn}
\usepackage{bm}
\usepackage{amsmath,amssymb}
\usepackage{hyperref}
\usepackage{natbib}
\usepackage{xcolor}
\usepackage[normalem]{ulem}
\usepackage{booktabs}
\usepackage{tabularx}

\usepackage[font=small,format=plain,labelfont=bf,
singlelinecheck=false,justification=raggedright,
labelsep=space,figurename=FIG.]{caption}

\begin{document}
	
\preprint{APS/123-QED}
	
\title{Tailoring magnetic interactions in atomic bilayers of Rh and Fe on Re(0001)}
	
\author{Souvik Paul}
\email{paul@physik.uni-kiel.de}
\author{Stefan Heinze}
	
\affiliation{Institute of Theoretical Physics and Astrophysics, Christian-Albrechts-Universit$\ddot{a}$t zu Kiel, Leibnizstrasse 15, 24098 Kiel, Germany}
	
	\date{\today}

\begin{abstract}
Using density functional theory, we investigate the interplay among the sequence of bilayers composed of an Fe and a Rh layer on the Re(0001), their stacking order and their magnetic properties. We find that bilayers in which both layers are in fcc stacking are energetically very unfavorable, while all other combinations of hcp and fcc stacking are energetically relatively close. The magnetic interactions are evaluated by mapping the DFT energies onto an atomistic spin model, which contains the Heisenberg exchange, the Dzyaloshinskii-Moriya interaction (DMI), the magnetocrystalline anisotropy energy, and higher-order exchange interactions. We find that the stacking sequence of the bilayer significantly modifies the magnetic interactions. For bilayers in which Fe is the topmost layer, the nearest-neighbor exchange interaction is ferromagnetic, but varies in strength by a factor of up to three for different stacking sequences. The DMI even changes up to a factor of four. As a result, we find a DMI driven cycloidal spin spiral ground state with a period of 11~nm for hcp-Fe/hcp-Rh/Re(0001). For fcc-Fe/hcp-Rh/Re(0001) and hcp-Fe/fcc-Rh/Re(0001), we obtain a ferromagnetic ground state. The spin spiral energy dispersion of hcp-Fe/hcp-Rh/Re(0001) including spin-orbit coupling suggests that isolated skyrmions can be stabilized in the field-polarized ferromagnetic background at external magnetic fields. If the Fe layer is sandwiched between the Rh overlayer and the Re(0001) substrate, there is a competition between the ferromagnetic coupling preferred by the Rh-Fe hybridization and the antiferromagnetic coupling induced by the Fe-Re hybridization. Due to the Fe/Re interface, the DMI can become very large. For fcc-Rh/hcp-Fe/Re(0001), we obtain a cycloidal spin spiral with a period of 1.7~nm, which is induced by frustration of exchange interactions and further stabilized by the DMI. For hcp-Rh/hcp-Fe/Re(0001), we find a DMI driven cycloidal spin spiral with a period of 4~nm and locally nearly antiparallel magnetic moments due to antiferromagnetic nearest-neighbor exchange. The higher-order exchange constants can be significant in the considered films, however, they do not stabilize multi-$Q$ states. 
\end{abstract}

\maketitle

\section{Introduction}

Driven by the demand for next generation energy-efficient, high-speed, high-density, and low-dimensional spintronic devices, magnetic nanostructures at surfaces, interfaces and in multilayers are currently a research focus in spintronics \cite{Hellman2017}. Transition-metal monolayers on metal surfaces are important model systems in this field since they allow to study the underlying microscopic mechanisms and magnetic interactions. For example, the interfacial Dzyaloshinskii-Moriya interaction \cite{dmi1,dmi2,dmi3}, which arises due to broken inversion symmetry and spin-orbit coupling, has been first observed in a Mn monolayer on the W(110) surface \cite{dmi4}. Subsequently, the discovery of a nanoskyrmion lattice in an Fe monolayer on Ir(111) \cite{tf1} opened the route towards exploring skyrmions at transition-metal interfaces and in multilayers \cite{rt1,tf2,ml1,ml2,ml3,ml4,Dupe2016}.

The potential of transition-metal interfaces is impressively revealed by the observation of isolated magnetic skyrmions in a Pd/Fe bilayer on the Ir(111) surface \cite{tf2,Romming2015}, i.e., already a single atomic layer of Pd can change the magnetic state drastically. Based on first-principles calculations, it has been explained that the transition from a skyrmion lattice to isolated skyrmions is due to the modification of the exchange interactions rather than the Dzyaloshinskii-Moriya interaction \cite{Dupe2014,Simon2014}. Recently, it has been demonstrated that a similar frustration of exchange interactions along with the Dzyaloshinskii-Moriya interaction can induced zero-field skyrmions in Rh/Co bilayers on the Ir(111) surface \cite{Meyer2019}. 

In contrast, a Rh/Fe bilayer on Ir(111) exhibits qualitatively different spin structures \cite{rhfeir111}. Both hcp and fcc stacking of the Rh overlayer have been observed, while Fe only prefers fcc stacking, implying that the ground state spin structure is determined by overlayer stacking \cite{rhfeir111}. For fcc-Rh stacking, a spin spiral with a wavelength of 1.5 nm is found which is driven by frustration of exchange interactions as explained from first-principles calculations and a skyrmion phase could not be obtained at experimentally feasible magnetic fields. For hcp-Rh stacking, it was shown that due to a competition of the higher-order exchange interactions and the Dzyaloshinskii-Moriya interaction (DMI), a canted double-row wise antiferromagnetic ($\uparrow \uparrow \downarrow \downarrow$) state occurs which was observed by spin-polarized scanning tunneling microscopy. The $\uparrow \uparrow \downarrow \downarrow$ state was also observed in an Fe monolayer on the Rh(111) surface \cite{3spin} and explained based on introducing a three-site four spin interaction \cite{hoffmann} in addition to the previously considered two-site four spin and four-site four spin interaction \cite{hubbard2}. 

Here, we address the question that how far the spin structures observed in Rh/Fe bilayers on Ir(111) and their origin of the strong higher-order exchange interactions depend on the transition-metal substrate and the stacking order of the bilayer. For this purpose, we have performed a systematic study using density functional theory (DFT) on the impact of the stacking sequences of atomic bilayers composed of Rh and Fe on the Re(0001) surface on the magnetic ground state. We consider both films with Fe at the surface, denoted as Fe/Rh bilayers, and films with Fe sandwiched between the Rh overlayer and the Re substrate, i.e., Rh/Fe bilayers. All four possible combinations of hcp and fcc stacking for the bilayers are taken into consideration. Our total energy calculations suggest that except for the fcc/fcc stacking all other six stacking orders could be realized in experiments. We also demonstrate that the magnetic interactions depend significantly on the stacking sequences of the Fe/Rh and Rh/Fe bilayers. As a consequence, we find noncollinear spin spiral ground states with different periods ranging from 1.7~nm to 11~nm in hcp-Fe/hcp-Rh/Re(0001), hcp-Rh/hcp-Fe/Re(0001) and fcc-Rh/hcp-Fe/Re(0001) and a ferromagnetic ground state in other systems. Interestingly, the energy dispersion of hcp-Fe/hcp-Rh/Re(0001) reveals that it can host isolated skyrmions in external magnetic fields. 

Based on our DFT calculations, we parametrize an atomistic spin model including the Heisenberg exchange interactions, the DMI, the magnetocrystalline anisotropy energy as well as the higher-order exchange interactions. The latter interactions are significant but do not determine the magnetic ground state as found for Rh/Fe bilayers on Ir(111). This shows that the substrate has a decisive influence on the properties of the bilayers. Since Fe monolayers have been pseudomorphically grown on Re(0001) experimentally \cite{fere0001e}
and the predicted noncollinear magnetic ground state \cite{hardrat} has been confirmed in those experiments, the ultrathin film systems proposed here seem promising for future experiments.

This paper is organized into three sections. The first section (Sec. \ref{sec:comdet}) contains the methodology used to calculate the structural and magnetic properties. In the second section (Sec. \ref{sec:resdis}), we present results regarding the different stacking orders of Fe/Rh/Re(0001) and Rh/Fe/Re(0001). We summarize our results in the last sections (Sec. \ref{sec:conc}).          

\section{\label{sec:comdet}Computational details}

The electronic structures of Fe/Rh and Rh/Fe bilayers on Re(0001) have been calculated using the full-potential linearized augmented plane wave (FP-LAPW) method based \textsc{fleur} code \cite{fleur} and the projected augmented wave method based \textsc{vasp} (Vienna Ab initio Simulation Package) code \cite{vasp}. The consistency and accuracy of these two DFT codes in calculating the necessary quantities are discussed in section \ref{sec:resdis}. The structural relaxed parameters and energy of the multi-$Q$ states were computed with the \textsc{vasp} code, whereas the energy dispersion of spin spirals without and with spin-orbit coupling (SOC) and magnetocrystalline anisotropy energy (MAE) were computed with the \textsc{fleur} code. The experimental lattice constants of bulk Re, $a= 2.761$ \AA\ and $c/2= 2.228$ \AA\ \cite{re}, were used as in-plane lattice constant ($a_{in}$) and interlayer distance ($d_{0}$), respectively, in film geometry. In the \textsc{fleur} code, muffin-tin radii of 2.3 a.u. (1.22 \AA) were used for Fe and Rh, and 2.45 a.u. (1.30 \AA) was chosen for Re. 

To find an optimum geometry of the ultrathin films, we used the generalized gradient approximation (GGA) parametrized by Perdew, Burke and Ernzerhof as the exchange-correlation (XC) part of the potential \cite{ggapbe}. The top three layers were relaxed along the $z$-direction until forces on all atoms were less than 0.04 eV/\AA\ in \textsc{fleur} and 0.01 eV/\AA\ in \textsc{vasp}. The Brillouin zone (BZ) integration was carried out on 66 $k$ points in the irreducible part of the two-dimensional BZ (2DBZ). The energy cut-off of the plane wave was chosen as $k_{max}= 4.0$ a.u.$^{-1}$ for \textsc{fleur} and 500 eV for \textsc{vasp}.

An efficient way of searching for a noncollinear magnetic ground state in ultrathin films is to calculate the energy dispersion of spin spirals \cite{kurz2004}. Then the energy of the spin spirals can be mapped onto the Heisenberg model for understanding the stability of noncollinear state via the magnetic exchange interactions. The Heisenberg model is given by

\begin{align} \label{eq1}
\mathcal{H} =- \sum_{ij} J_{ij} (\textbf{m}_{i}\cdot\textbf{m}_{j})
\end{align}
where $J_{ij}$ is the exchange constant and $\textbf{m}_{i}$ ($\textbf{m}_{j}$) is the unit magnetization vector at site $i$ ($j$).

The energy of homogeneous flat spin spirals was calculated using the \textsc{fleur} code as a function of wave vector \textbf{q} along two high symmetric directions of 2DBZ. To avoid supercell calculations and reduce computational time, we used the generalized Bloch theorem \cite{gbt} to compute the spin spiral energy within the chemical unit cell \cite{kurz2004}. We modeled the surface by an asymmetric film consisting of the Fe/Rh or Rh/Fe bilayer on nine layers of Re(0001) substrate. We used a local density approximation (LDA) form given by Vosko, Wilk and Nusair \cite{vwn} in the XC functional and a dense mesh of 44$\times$44 $k$-points in the full 2DBZ to calculate the energy dispersion of spin spirals. The energy cutoff for the plane wave is set to $k_{max}=$ 4.0 a.u.$^{-1}$. For more accurate energy calculations around the $\overline{\Gamma}$ point [$\overline{\mathrm{M}}$ point for hcp-Rh/hcp-Fe/Re(0001)], i.e., $\mid\textbf{q}\mid \lesssim 0.1(\frac{2\pi}{a}$), a $k$-mesh of 48$\times$48 points and $k_{max}=$ 4.3 a.u.$^{-1}$ were used.

The energy degeneracy of left and right rotating spin spirals, as predicted by the Heisenberg model, is lifted when SOC is taken into account. SOC introduces two additional energy contributions, namely, the MAE and DMI. The latter arises due to broken inversion symmetry at the interface. The DMI Hamiltonian can be written as

\begin{align} \label{eq2}
\mathcal{H}_{DMI} =- \sum_{ij} \textbf{D}_{ij}\cdot(\textbf{m}_{i}\times\textbf{m}_{j})
\end{align}
here $\textbf{D}_{ij}$ is the DMI vector the magnitude of which determines the strength of this interaction. Due to symmetry of the ultrathin film, the DMI vectors lie in the surface plane
which promotes cycloidal spin spirals. 
The DMI energy is computed in first-order perturbation theory based on self-consistent spin spiral states \cite{heide}.

To evaluate the MAE, we have performed fully self-consistent relativistic (SOC) calculations in the second variational approach with in-plane ($\parallel$) and out-of-plane ($\perp$) magnetization directions. The MAE is defined as $K_{\mathrm{MAE}}$= $E_{\parallel}$ $-$ $E_{\perp}$, i.e., a positive (negative) value of $K_{\mathrm{MAE}}$ indicates an out-of-plane (in-plane) easy magnetization axis. We have used asymmetric films containing 13 to 17 substrate layers to obtain an accurate value of the MAE. 

The Heisenberg model can be obtained from the Hubbard model as a second-order perturbative expansion in the hopping parameter over the Coulomb interaction \cite{hubbard1,hubbard2}. The fourth-order perturbative expansion results in a two-site four spin, a three-site four spin \cite{hubbard2,hubbard3} and a four-site four spin \cite{hoffmann} interaction, which can be understood as hopping of electrons among two, three and four lattice sites, respectively. For simplicity, we denote these terms as biquadratic, 3-spin, and 4-spin interactions, respectively and they are as follows

\begin{gather}
\mathcal{H}_{\mathrm{biquad}} = - \sum_{ij} B_{ij} (\textbf{m}_{i} \cdot \textbf{m}_{j})^2 \\
\mathcal{H}_{\mathrm{3-spin}} = - 2 \sum_{ijk} Y_{ijk} (\textbf{m}_{i} \cdot \textbf{m}_{j}) (\textbf{m}_{j} \cdot \textbf{m}_{k}) \\
\mathcal{H}_{\mathrm{4-spin}} = - \sum_{ijkl} K_{ijkl} [(\textbf{m}_{i} \cdot \textbf{m}_{j}) (\textbf{m}_{k} \cdot \textbf{m}_{l}) \nonumber \\
+(\textbf{m}_{i} \cdot \textbf{m}_{l}) (\textbf{m}_{j} \cdot \textbf{m}_{k})-(\textbf{m}_{i} \cdot \textbf{m}_{k}) (\textbf{m}_{j} \cdot \textbf{m}_{l})]
\end{gather}

\begin{table*}[!htbp]
	\centering
	\caption{Structural relaxed parameters for four stacking sequences of Fe/Rh bilayer on Re(0001). The interlayer distance and magnetic moments of the three outermost layers for the relaxed geometry in the FM state are calculated using $\textsc{vasp}$ ($\textsc{Fleur}$). The energy difference between the AFM and FM states ($E_\mathrm{{AFM}}-E_\mathrm{{FM}}$) and energy of the FM states ($E_{\mathrm{relative}}^\mathrm{{FM}}$) relative to the lowest FM state are also shown.}
	\label{tab:table1}
	\begin{ruledtabular}
		\begin{tabular}{ccccccccc} 
			Stacking & \begin{tabular}[c]{@{}c@{}}$\Delta d_{\mathrm{Fe-Rh}}$\\ (\%)\end{tabular} & \begin{tabular}[c]{@{}c@{}}$\Delta d_{\mathrm{Rh-Re}}$\\ (\%)\end{tabular} & \begin{tabular}[c]{@{}c@{}}$\Delta d_{\mathrm{Re-Re}}$\\ (\%)\end{tabular} & $\mu_{\mathrm{Fe}}$ & $\mu_{\mathrm{Rh}}$ & $\mu_{\mathrm{Re}}$ & \begin{tabular}[c]{@{}c@{}}$E_\mathrm{{AFM}}-E_\mathrm{{FM}}$\\ (meV/Fe)\end{tabular} & \begin{tabular}[c]{@{}c@{}}$E_{\mathrm{relative}}^\mathrm{{FM}}$\\ (meV/Fe)\end{tabular} \\
			\colrule
			fcc-Fe/hcp-Rh & \begin{tabular}[c]{@{}c@{}}$-$4.40\\ ($-$4.53)\end{tabular} & \begin{tabular}[c]{@{}c@{}}$-$2.15\\ ($-$1.80)\end{tabular} & \begin{tabular}[c]{@{}c@{}}$-$0.81\\ ($-$0.36)\end{tabular} & \begin{tabular}[c]{@{}c@{}}2.93\\ (2.96)\end{tabular} & \begin{tabular}[c]{@{}c@{}}0.18\\ (0.13)\end{tabular} & \begin{tabular}[c]{@{}c@{}}$-$0.08\\ ($-$0.07)\end{tabular} & \begin{tabular}[c]{@{}c@{}}98.6\\ (115.2)\end{tabular} & \begin{tabular}[c]{@{}c@{}}33.7\\ (23.2)\end{tabular}\\
			hcp-Fe/hcp-Rh & \begin{tabular}[c]{@{}c@{}}$-$5.75\\ ($-$5.47)\end{tabular} & \begin{tabular}[c]{@{}c@{}}$-$3.05\\ ($-$2.60)\end{tabular} & \begin{tabular}[c]{@{}c@{}}$-$0.36\\ ($-$0.22)\end{tabular} & \begin{tabular}[c]{@{}c@{}}2.90\\ (2.84)\end{tabular} & \begin{tabular}[c]{@{}c@{}}0.11\\ (0.13)\end{tabular} & \begin{tabular}[c]{@{}c@{}}$-$0.08\\ ($-$0.06)\end{tabular} & \begin{tabular}[c]{@{}c@{}}61.7\\ (78.6)\end{tabular} & \begin{tabular}[c]{@{}c@{}}0.0\\ (0.0)\end{tabular}\\
			hcp-Fe/fcc-Rh & \begin{tabular}[c]{@{}c@{}}$-$7.54\\ ($-$6.69)\end{tabular} & \begin{tabular}[c]{@{}c@{}}$-$1.26\\ ($-$1.00)\end{tabular} & \begin{tabular}[c]{@{}c@{}}1.44\\ (2.60)\end{tabular} & \begin{tabular}[c]{@{}c@{}}2.87\\ (2.91)\end{tabular} & \begin{tabular}[c]{@{}c@{}}0.13\\ (0.12)\end{tabular} & \begin{tabular}[c]{@{}c@{}}$-$0.08\\ ($-$0.08)\end{tabular} & \begin{tabular}[c]{@{}c@{}}21.9\\ (29.9)\end{tabular} & \begin{tabular}[c]{@{}c@{}}124.7\\ (102.9)\end{tabular}\\
			fcc-Fe/fcc-Rh & 6.37 & $-$1.71 & 1.44 & 2.83 & 0.10 & $-$0.09 & 48.5 & 650.2 \\
		\end{tabular}
	\end{ruledtabular}
\end{table*}

In this paper, we only consider the nearest-neighbor exchange constant of these three high-order (HO) terms, i.e., B$_{1}$, Y$_{1}$ and K$_{1}$. We evaluated these constants from energies of three multi-$Q$ states: (i) two two-dimensionally modulated collinear spin structures along $\overline{\Gamma \mathrm{K}}$ and $\overline{\Gamma\mathrm{M}}$, the so-called up-up-down-down ($uudd$) 
or double-row wise antiferromagnetic state \cite{hardrat} and (ii) a three-dimensionally modulated noncollinear spin structure at the $\overline{\mathrm{M}}$ point, the so-called 3$Q$ state \cite{pkurz}. The multi-$Q$ states can be constructed from the superposition of spin spirals corresponding to the symmetry equivalent $\textbf{q}$ vectors of 2DBZ. By construction, they are degenerate with the spin spiral states (single-$Q$ states) in the Heisenberg model. However, within the extended-Heisenberg model, the degeneracy is lifted and the energy differences between the multi-$Q$ and single-$Q$ states, as obtained from DFT calculations, are connected to the HO exchange constants as follows \cite{hoffmann}

\begin{gather} 
B_{1}=\frac{3}{32} \Delta E_{\overline{\mathrm{M}}}^{3Q} - \frac{1}{8} \Delta E_{\overline{\mathrm{M}}/2}^{uudd} \label{eq6} \\
Y_{1}=\frac{1}{8}(\Delta E_{3\overline{\mathrm{K}}/4}^{uudd} - \Delta E_{\overline{\mathrm{M}}/2}^{uudd}) \label{eq7} \\
K_{1}=\frac{3}{64} \Delta E_{\overline{\mathrm{M}}}^{3Q} + \frac{1}{16} \Delta E_{3{\overline{\mathrm{K}}}/4}^{uudd} \label{eq8}
\end{gather}
where the $\Delta E$s are
\begin{gather}
\Delta E_{\overline{\mathrm{M}}}^{3Q}= E_{\overline{\mathrm{M}}}^{3Q} - E_{\overline{\mathrm{M}}}^{SS}	 \label{eq9} \\
\Delta E_{\overline{\mathrm{M}}/2}^{uudd}= E_{\overline{\mathrm{M}}/2}^{uudd} - E_{\overline{\mathrm{M}}/2}^{SS}  \label{eq10} \\
\Delta E_{3\overline{\mathrm{K}}/4}^{uudd}= E_{3{\overline{\mathrm{K}}}/4}^{uudd} - E_{3{\overline{\mathrm{K}}}/4}^{SS} \label{eq11} \\
\nonumber
\end{gather}

Here $\Delta E_{\overline{\mathrm{M}}}^{3Q}$ is the energy difference between the 3$Q$ and spin spiral state at the $\overline{\mathrm{M}}$ point along $\overline{\Gamma \mathrm{KM}}$, $\Delta E_{\overline{\mathrm{M}}/2}^{uudd}$ is the energy difference between the $uudd$ and the spin spiral state at $\overline{\mathrm{M}}/2$ along $\overline{\Gamma \mathrm{M}}$ and $\Delta E_{3\overline{\mathrm{K}}/4}^{uudd}$ is the energy difference between the $uudd$ and spin spiral state at $3\overline{\mathrm{K}}/4$ along $\overline{\Gamma \mathrm{K}}$. 
 
\section{\label{sec:resdis}Results and discussions}

\subsection{\label{sec:colmag}Collinear magnetic states in Fe/Rh bilayer on Re(0001)}

We begin our investigation by presenting results on collinear magnetic configurations [FM and $c$(2$\times$2) AFM states] of Fe/Rh bilayers on Re(0001) as computed by $\textsc{vasp}$ and $\textsc{fleur}$ in Table \ref{tab:table1}. Four possible stacking sequences of Fe/Rh bilayers on Re(0001) are examined: fcc-Fe/hcp-Rh, hcp-Fe/hcp-Rh, hcp-Fe/fcc-Fe and fcc-Fe/hcp-Rh. Hereafter, for brevity in writing, we drop the substrate part.    

To obtain an optimum interlayer distance and collinear magnetic ground state, we first relax the three topmost layers of the ultrathin films along the out-of-plane direction in the FM and AFM states, keeping the other substrate layers fixed. The calculated energy difference $\Delta E$=$E_{\mathrm{AFM}}-E_{\mathrm{FM}}$ demonstrates that the ground state prefers a ferromagnetic (FM) alignment for all the films. The FM state of hcp-Fe/hcp-Rh has the lowest energy followed by fcc-Fe/hcp-Rh and hcp-Fe/fcc-Rh, whereas fcc-Fe/fcc-Rh has the highest energy (650 meV/Fe). Since the relative energy of fcc-Fe/hcp-Rh and hcp-Fe/fcc-Rh films are not too high ($\sim$100 meV/Fe), these two along with hcp-Fe/hcp-Rh can appear in the pseudomorphic growth of experimental film preparation. 

The relaxed interlayer distance and magnetic moments of the ground state are listed in Table \ref{tab:table1}. The relaxed distance between layers a and b is expressed as

\begin{gather}
\Delta d_{\mathrm{a-b}} (\%)= (\frac{d_{\mathrm{ab}}-d_{0}}{d_{0}}) \times 100 \label{eq12}
\end{gather}
where $d_{\mathrm{ab}}$ is the relaxed interlayer distance and $d_{0}$ is the bulk interlayer distance of Re (2.228 \AA). In general, due to lower coordination number compared to Rh and Ir, Fe shows a sizable amount of relaxation. For the first three cases (Table \ref{tab:table1}), the topmost Fe layer moves closer to the Rh layer and the interlayer spacing ($\Delta d_{\mathrm{Fe-Rh}}$) varies from $-4.4\%$ to $-7.5\%$. However, for fcc-Fe/fcc-Rh, the Fe-Rh interlayer relaxation is outward ($+6\%$), which indicates that the symmetry of the bilayer can be a decisive factor. For all the four cases, the second interlayer relaxation ($\Delta d_{\mathrm{Rh-Re}}$) is inward and the values are quite small compared to the first one ($\Delta d_{\mathrm{Fe-Rh}}$). The first substrate interlayer relaxation ($\Delta d_{\mathrm{Re-Re}}$) is small and depends on the symmetry of the Rh layer. For hcp-stacked Rh films, i.e., fcc-Fe/hcp-Rh and hcp-Fe/hcp-Rh, the relaxation is inward and the values are an order of magnitude smaller than $\Delta d_{\mathrm{Rh-Re}}$. On the other hand, for hcp-Fe/fcc-Rh and fcc-Fe/fcc-Rh, $\Delta d_{\mathrm{Re-Re}}$ is positive (outward relaxation) and the values have same order of magnitude as $\Delta d_{\mathrm{Rh-Re}}$. 

We notice that the ferromagnetic Fe layer induces magnetic moments on Rh and the first Re substrate layer which are aligned parallel and antiparallel with the Fe moments, respectively. The induced moments on Rh are very small ($\sim 0.1 \mu_{B}$) and on Re are even smaller. A close inspection reveals that as the moments of the Fe layer grow, the interlayer distance $\Delta d_{\mathrm{Fe-Rh}}$ becomes smaller, which indicates that the larger magnetic moments are caused by smaller relaxation. Among the structural data obtained using $\textsc{vasp}$, the Fe layer exhibits the largest inward relaxation for hcp-Fe/fcc-Rh, the smallest relaxation for fcc-Fe/hcp-Rh and a intermediate one for hcp-Fe/hcp-Rh. 

The relaxed interlayer distances, the magnetic moments, the energies calculated for the first three films in Table \ref{tab:table1} using $\textsc{fleur}$ match quite well with the $\textsc{vasp}$ results. This permit us to conduct comparative analyses of properties using different implementations of first-principles codes ($\textsc{vasp}$ and $\textsc{fleur}$). Since the total energy of fcc-Fe/fcc-Rh is very high compared to all other films, we exclude this system from further discussions.

\subsection{\label{sec:ncolmag}Noncollinear magnetism in Fe/Rh bilayer on Re(0001)}

\begin{figure}[!h]
	\includegraphics[scale=1.0]{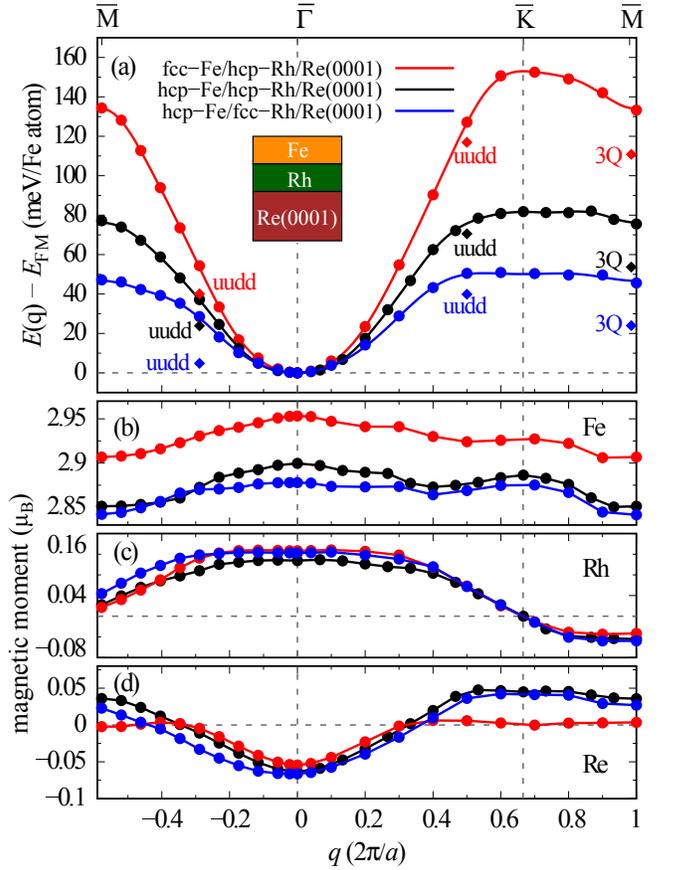}
	\centering
	\caption{(a) Energy dispersion $E(\mathbf{q})$ of flat spin spirals along two high symmetry directions ($\overline{\Gamma \mathrm{KM}}$ and $\overline{\Gamma \mathrm{M}}$) for fcc-Fe/hcp-Rh/Re(0001) (red), hcp-Fe/hcp-Rh/Re(0001) (black) and hcp-Fe/fcc-Rh/Re(0001) (blue). The filled circles represent DFT data and the solid lines are the fit to the Heisenberg model. The filled diamonds represent the 3$Q$ and $uudd$ states at the $\textbf{q}$ points corresponding to the single-$Q$ states. Magnetic moments of Fe, Rh and Re (third layer) are shown in panel (b), (c) and (d), respectively.} 
	\label{fig:f1}
\end{figure}

\begin{table*}[!thbp]
	\centering
	\caption{Effective exchange constant ($J_{\mathrm{eff}}$), exchange parameters ($J_{ij}$), biquadratic exchange constant ($B_{1}$), 3-site four spin exchange constant ($Y_{1}$) and 4-site four spin exchange constant ($K_{1}$) for three stackings of Fe/Rh bilayer on Re(0001). $\Delta E$ is the energy difference between the multi-$Q$ and spin spiral states as defined by Eqs.~(9-11). All energies are given in meV.} 
	\label{tab:table2}
	\begin{ruledtabular}
		\begin{tabular}{ccccccccccccccc}
			Stacking & $J_{\mathrm{eff}}$ & $J_{1}$ & $J_{2}$ & $J_{3}$ & $J_{4}$ & $J_{5}$ & $J_{6}$ & $J_{7}$ & $B_{1}$ & $Y_{1}$ & $K_{1}$ & $\Delta E_{\overline{\mathrm{M}}}^{3Q}$ & $\Delta E_{3\overline{\mathrm{K}}/4}^{uudd}$ & $\Delta E_{\overline{\mathrm{M}}/2}^{uudd}$\\
			\colrule
			fcc-Fe/hcp-Rh & 9.62 & 18.25 & $-1.79$ & $-1.40$ & 0.41 & 0.09 & 0.10 & $-$0.41 & $-$0.33 & 0.49 & $-$1.69 & $-$22.42 & $-$10.26 &$-$14.21\\
			hcp-Fe/hcp-Rh & 6.36 & 9.85 & 0.23 & $-$0.15 & $-$0.22 & 0.27 & 0.05 & $-$0.16 & $-$0.39 & 1.00 & $-$1.36 & $-$21.88 & $-$5.31 & $-$13.33\\
			hcp-Fe/fcc-Rh & 5.43 & 5.73 & 0.50 & 0.47 & $-$0.26 & 0.12 & 0.02 & $-$0.07 & 0.94 & 1.64 & $-$1.68 & $-$21.67 & $-$10.65 & $-$23.77\\
		\end{tabular}
	\end{ruledtabular}
\end{table*}

In the previous section, we have investigated collinear magnetic states and found that the FM configuration is preferred over the AFM one. In this section, we first study the possibility of a noncollinear ground state without SOC. Later, the DMI and MAE contributions are added to the spin spiral energy dispersion to find the true ground state. We also present the energies of the multi-$Q$ states.  

In Fig.~\ref{fig:f1}(a), we present the energy dispersion of homogeneous flat spin spirals as a function of wave vector $\textbf{q}$, calculated using $\textsc{fleur}$, along two high symmetry directions $\overline{\Gamma \mathrm{KM}}$ and $\overline{\Gamma \mathrm{M}}$ of the 2DBZ for three films: fcc-Fe/hcp-Rh, hcp-Fe/hcp-Rh and hcp-Fe/fcc-Rh on Re(0001). The energy dispersion curves suggest that the magnetic ground state is ferromagnetic, which does not modify our conclusion from collinear calculations. The energy difference between the FM state (at the $\overline{\Gamma}$ point) and the AFM state (at the $\overline{\mathrm{M}}$ point), $\Delta E$=$E_\mathrm{{AFM}}-E_\mathrm{{FM}}$, is 133 meV/Fe in fcc-Fe/hcp-Rh and the difference reduces to 45 meV/Fe in hcp-Fe/fcc-Rh, which result in trimming the slope of the dispersion curve at the $\overline{\Gamma}$ point. The values of $\Delta E$ in Fig.~\ref{fig:f1}(a) are slightly larger compared to the values in Table \ref{tab:table1}. The reasons are: (i) the energies of Table \ref{tab:table1} are calculated with the GGA functional, while that of Fig.~\ref{fig:f1}(a) are calculated with the LDA functional and (ii) the total energies of the collinear states in Table \ref{tab:table1} are computed based on the relaxed parameters obtained for the respective magnetic state, whereas the relaxed interlayers of the FM state are used to compute the total energies of both states in Fig.~\ref{fig:f1}.  

To examine the position of multi-$Q$ states in the energy landscape, we compute the energy of the $uudd$ states along $\overline{\Gamma \mathrm{K}}$ and along $\overline{\Gamma \mathrm{M}}$ as well as the $3Q$ state. In Fig.~\ref{fig:f1}(a), these states are designated by diamonds. The $uudd$ state along $\overline{\Gamma \mathrm{KM}}$ and the $3Q$ state are quite high in energy compared to the ferromagnetic state, while the $uudd$ state along $\overline{\Gamma \mathrm{M}}$ is relatively closer. Interestingly, the $uudd$ state of hcp-Fe/fcc-Rh along $\overline{\Gamma \mathrm{M}}$ is only 4.8 meV/Fe higher than the ground state. With a small perturbation to the collinear angle, this $uudd$ state can lose the surplus energy compared to the FM state and may become the ground state of hcp-Fe/fcc-Rh as observed in hcp-Rh/Fe/Ir(111) \cite{rhfeir111}.          

In Fig.~\ref{fig:f1}(b-d), we present the magnetic moments of Fe, Rh and Re atoms, respectively. A strong spin-polarization of Fe induces small magnetic moments on Rh and nearly negligible amount on Re. The variation of the Fe magnetic moments with $\textbf{q}$ is fairly constant for the three films, which suggests that its magnetic interactions can be described within the Heisenberg model. We exploit this condition to map the total energies of spin spirals onto the Heisenberg Hamiltonian [Eq. (\ref{eq1})] to extract the exchange parameters. Note that the magnetic moments of Fe follow the trend in $\Delta E$.  

In Table \ref{tab:table2}, we list the seven exchange constants for all three systems, which are used in fitting. In addition, we also list the effective exchange constant, obtained from a fit in the vicinity ($\mid$$\textbf{q}$$\mid \leq 0.1\times\frac{2\pi}{a}$) of $\overline{\Gamma}$. The effective exchange indicates the behavior of the curve within a very small region around $\overline{\Gamma}$. The positive value of the effective exchange constants suggest that the ground of the three films is ferromagnetic. These constants also specify that the slopes of hcp-Fe/hcp-Rh and hcp-Fe/fcc-Rh are close to each other, while fcc-Fe/hcp-Rh has a significantly larger slope. All our systems are dominated by the nearest-neighbor exchange interaction, which follows the slope of the dispersion curve for $\mid$$\textbf{q}$$\mid \geq 0.2\times\frac{2\pi}{a}$. We see that some exchange constants are negative, however, due to their small magnitude, they fail to introduce a significant amount of frustration into the system and cannot stabilize a spin spiral state. 

The energy differences of the three multi-$Q$ states with respect to the corresponding spin spiral states are listed in Table \ref{tab:table2}. All the multi-$Q$ states are energetically lower than the corresponding spin spiral states. The HO exchange constants are obtained by solving Eqs.~(\ref{eq6}-\ref{eq8}). The 4-site four spin constants ($K_{1}$) have nearly the same value for the three systems, whereas the biquadratic ($B_{1}$) and 3-site four spin ($Y_{1}$) parameters vary by an order of magnitude depending on the stacking of the Fe/Rh bilayer. The 3-site four spin constant is $\sim$0.5 meV for fcc-Fe/hcp-Rh and it increases to 1 meV (1.6 meV) for hcp-Fe/hcp-Rh (hcp-Fe/fcc-Rh). Likewise, the biquadratic constant is $\sim$$-$0.3 meV for fcc-Fe/hcp-Rh and hcp-Fe/hcp-Rh and it becomes $\sim$0.9 meV for hcp-Fe/fcc-Rh. Before explaining the trends in the HO exchange constants, note that out of the three energy difference, $\Delta E_{\overline{\mathrm{M}}}^{3Q}$, $\Delta E_{\overline{\mathrm{M}}/2}^{uudd}$ and $\Delta E_{3\overline{\mathrm{K}}/2}^{uudd}$, only two are required to compute a particular higher-order constant [Eqs. (\ref{eq6}-\ref{eq8})]. The energy of the 3$Q$ state with respect to the spin spiral state, $\Delta E_{\overline{\mathrm{M}}}^{3Q}$, almost remains unchanged for the three films. Therefore, the same energy difference between the $uudd$ and $\textbf{q}$=$3\overline{\mathrm{K}}/4$ states in fcc-Fe/hcp-Rh and hcp-Fe/fcc-Rh ($\Delta E_{3\overline{\mathrm{K}}/4}^{uudd}$$\approx$ $-$10 meV) leads to a same 4-site four spin constant ($-$1.7 meV) according to Eq. (\ref{eq8}). The 4-site four spin constant is reduced slightly in hcp-Fe/hcp-Rh ($-$1.4 meV) as the energy difference ($\Delta E_{3\overline{\mathrm{K}}/4}^{uudd}$$\approx$ $-$5 meV) is reduced by a factor of two. By considering that the values of $\Delta E_{\overline{\mathrm{M}}}^{3Q}$ are constant for three films, it is trivial to comprehend the trend of the biquadratic constants from Eq. (\ref{eq6}) using $\Delta E_{\overline{\mathrm{M}}/2}^{uudd}$. The 3-site four spin constants reflect the difference between $\Delta E_{3\overline{\mathrm{K}}/4}^{uudd}$ and $\Delta E_{\overline{\mathrm{M}}/2}^{uudd}$, but scaled by a factor of eight as depicted in Eq. (\ref{eq7}).              

\begin{table} [!htbp]
	\centering
	\caption{Effective Dzyaloshinskii-Moriya interaction constant ($D_{\mathrm{eff}}$), magnetocrystalline anisotropy ($K_{\mathrm{MAE}}$) for three stackings of Fe/Rh bilayer on Re(0001). The positive (negative) value of $K_{\mathrm{MAE}}$ indicates an out-of-plane (in-plane) easy magnetization axis. All energies are given in meV.}
	\label{tab:table3}
	\begin{ruledtabular}
		\begin{tabular}{ccc}
			Stacking & $D_{\mathrm{eff}}$ & $K_{\mathrm{MAE}}$  \\
			\colrule
			fcc-Fe/hcp-Rh & 0.22 & 0.30  \\
			hcp-Fe/hcp-Rh & 0.89 & $-$0.20 \\
			hcp-Fe/fcc-Rh & 0.41 & $-$0.90 \\
		\end{tabular}
	\end{ruledtabular}
\end{table}

\begin{figure}[h!]
	\includegraphics[scale=0.98]{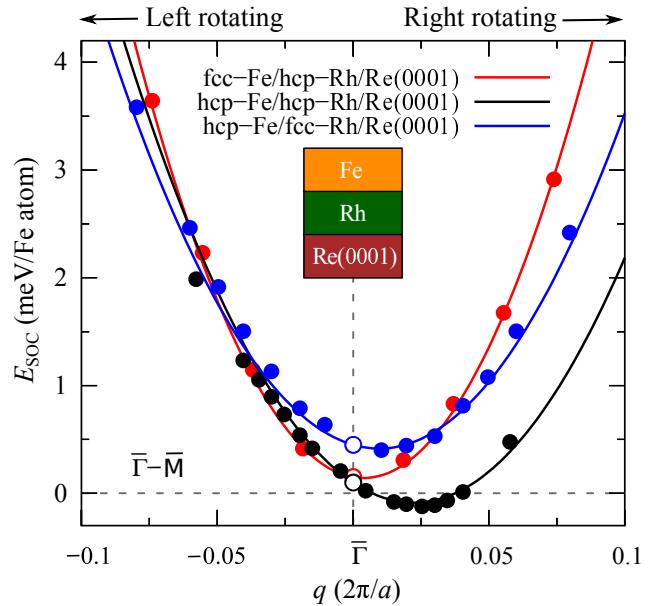} 
	\centering
	\caption{Energy dispersions of flat cycloidal spin spirals including DMI and MAE along $\overline{\Gamma \mathrm{M}}$ for fcc-Fe/hcp-Rh/Re(0001) (red), hcp-Fe/hcp-Rh/Re(0001) (black) and hcp-Fe/fcc-Rh/Re(0001) (blue). The filled circles represent DFT data and the solid lines are fits to the Heisenberg [Eq. (\ref{eq1})] and the DMI [Eq. (\ref{eq2})] models. MAE shifts the energy by $K_{\mathrm{MAE}}$/2 with respect to the FM state at the $\overline{\Gamma}$ point.} 
	\label{fig:f2}
\end{figure}

We have seen that none of the spin spiral states are favorable compared to the FM state. However, the situation can change and spin spiral states may be stabilized by DMI. Now, we discuss the effect of SOC on the spin spiral energy dispersion. When SOC is included, we have two additional contributions: DMI and MAE.

The MAE of the three films is summarized in Table \ref{tab:table3}. For fcc-Fe/hcp-Rh, we find that the easy axis is out-of-plane, whereas it is in-plane for the other two films. The computed value of MAE is 0.30 meV and 0.20 meV for fcc-Fe/hcp-Rh and hcp-Fe/hcp-Rh, respectively, whereas hcp-Fe/fcc-Rh has the largest MAE constant among the three ($K_{\mathrm{MAE}}$=0.90 meV). We have also calculated the MAE in the $y$ direction for hcp-Fe/hcp-Rh and hcp-Fe/fcc-Rh (data not shown) and the values are the same as in the $x$ direction. We shift the spin spiral energies relative to the FM state by $+\frac{1}{2} K_{\mathrm{MAE}}$ as the MAE disfavor all spin spirals equally at small $\textbf{q}$. 

In Fig. \ref{fig:f2}, we show the DMI and MAE contributions to the energy dispersion of spin spirals along $\overline{\Gamma \mathrm{M}}$ direction for the three films. The effective DMI constant is evaluated by fitting the SOC energies in the vicinity ($\mid$$\textbf{q}$$\mid \leq 0.1\times\frac{2\pi}{a}$) of $\overline{\Gamma}$ (Table \ref{tab:table3}). The positive DMI constant for all three films indicates that right-rotating (clockwise-rotating) spin spirals are favored. The DMI constant of fcc-Fe/hcp-Rh attains the smallest value which results in the most symmetric curve with respect to chirality (clockwise and anticlockwise rotation). As the DMI constant increases, the curve becomes more asymmetric. Overall, the ground state of fcc-Fe/hcp-Rh and hcp-Fe/fcc-Rh remains ferromagnetic upon including the contributions of SOC. However, a spin spiral state has becomes energetically favorable than the FM state in hcp-Fe/hcp-Rh. The DMI favors the spin spirals by moving the energy minimum to $q=0.025(\frac{2\pi}{a}$) ($\lambda$= 11.04 nm) which is $-0.2$ meV/Fe atom lower than the FM state.

The shallow noncollinear minimum in the dispersion profile of hcp-Fe/hcp-Rh/Re(0001) close to the FM state implies that this system can host isolated skyrmions with external magnetic field. In this ultrathin film, the MAE is in-plane which disfavors skyrmion formation. However, an external out-of-plane magnetic field of $\sim$1 T can compensate the in-plane MAE energy and therefore, isolated skyrmions can be (meta)stabilized in the FM background. Lately, several surveys reported metastable skyrmions with in-plane anisotropy \cite{huang12,sumilan14,lin15,mark16,leonov17.96,hayami19.99}. Using atomistic spin dynamics simulations with the interaction parameters obtained here, we observe (meta)stable isolated skyrmions at small magnetic field in hcp-Fe/hcp-Rh~\cite{mine}. Here, we do not see any noteworthy effect of the higher-order terms in determining the magnetic ground state. However, they play a significant role in the stability of isolated skyrmions. A detailed investigation of the isolated skyrmions in hcp-Fe/hcp-Rh/Re(0001) including the effect of HO exchange interactions is presented in Ref.~\cite{mine}.         
 
\subsection{\label{sec:estr} Trends of the spin spiral curves from electronic structures} 

\begin{figure}[!htbp]
	\includegraphics[scale=1.00]{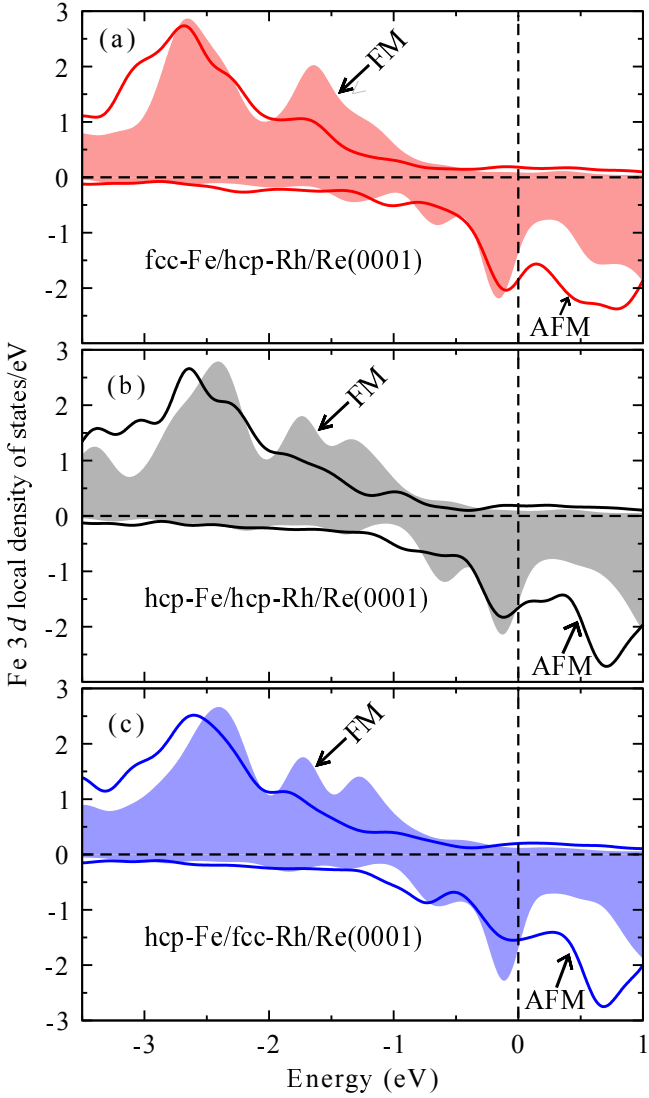}
	\centering
	\caption{Spin-polarized Fe 3$d$ local density of states (LDOS) for (a) fcc-Fe/hcp-Rh/Re(0001) (red), (b) hcp-Fe/hcp-Fe/Re(0001) (black) and (c) hcp-Fe/fcc-Rh/Re(0001) (blue). FM (filled) and AFM (line) LDOS are shown in each panel.} 
	\label{fig:f3}
\end{figure}

\begin{figure}[!htbp]
	\includegraphics[scale=1.0]{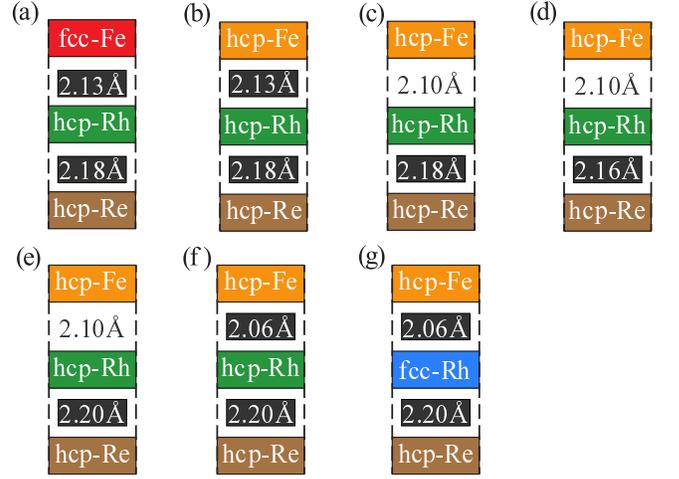}
	\centering
	\caption{Fe/Rh/Re(0001) films with various interlayer distances and stacking sequences. First three layers are shown for convenience. Films (a), (d) and (g) represent a relaxed geometry, i.e., their interlayer distances along with the order of Fe/Rh bilayer are consistent with Table \ref{tab:table1}. We introduce two intermediate films between (a)-(d) and (d)-(g) to study the change of $\Delta E$=$E_\mathrm{{AFM}}-E_\mathrm{{FM}}$. Each modification in the interlayer distance and stacking order is denoted by a change of color.} 
	\label{fig:f4}
\end{figure}

\begin{figure}[!htbp]
	\includegraphics[scale=1.0]{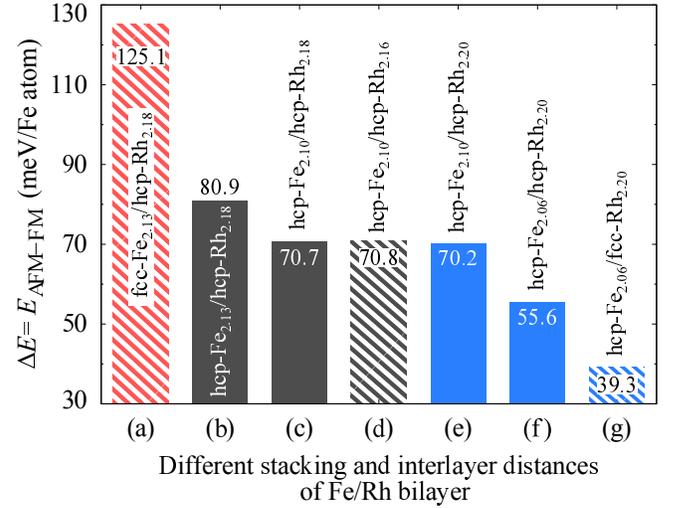}
	\centering
	\caption{Energy difference between the AFM and FM states ($\Delta E$) for films of Fig. \ref{fig:f4}. Films with relaxed geometry [(a), (d) and (g)] are shown using shaded color and intermediate films [(b), (c), (e) and (f)] are shown using solid color.} 
	\label{fig:f5}
\end{figure} 

In Fig.~\ref{fig:f1}(a), we have seen that the spin spiral dispersion curve becomes less steep as one moves through fcc-Fe/hcp-Rh, hcp-Fe/hcp-Rh and hcp-Fe/fcc-Rh. The quantity $\Delta E $=E$_{\mathrm{AFM}}-$E$_{\mathrm{FM}}$, i.e., the energy difference between the FM and AFM states, mimics the slope of the dispersion curves quite well. In this section, we intend to gain more insight into the trend of $\Delta E$ based on the electronic structure. Therefore, we present the local density of states (LDOS) in the FM and AFM configurations for the three films using $\textsc{vasp}$ in Fig.~\ref{fig:f3}. Each panel contains the Fe 3$d$ LDOS in the FM and AFM configurations. For consistency, we present the LDOS of one Fe atom in the AFM state, the other Fe atom has exactly the opposite LDOS.

First we compare the LDOS of fcc-Fe/hcp-Rh and hcp-Fe/hcp-Rh [Fig.~\ref{fig:f3}(a) and Fig.~\ref{fig:f3}(b)]. The broad majority peaks of the FM and AFM states around $-$2.6 eV for fcc-Fe/hcp-Rh in Fig. \ref{fig:f3}(a) change their position in Fig. \ref{fig:f3}(b) for hcp-Fe/hcp-Rh. The AFM peak moves to the lower energy and the FM peak moves towards the Fermi level. This opposite movement of the FM and AFM peaks reduces $\Delta E$ in hcp-Fe/hcp-Rh as compared to fcc-Fe/hcp-Rh. The majority FM peak at $-$1.6 eV in Fig.~\ref{fig:f3}(a) breaks into two peaks in Fig.~\ref{fig:f3}(b). One part moves to higher energy and the other one towards lower energy. This process, effectively, confers a very little contribution to $\Delta E$. The intensity of the minority FM peak, just below the Fermi level, remains same for fcc-Fe/hcp-Rh and hcp-Fe/hcp-Rh, but a relatively low intensity AFM peak in hcp-Fe/hcp-Rh further lowers $\Delta E$. A comparison of the LDOSs between Fig. \ref{fig:f3}(b) and Fig. \ref{fig:f3}(c) reveals that the intensity of the minority FM peak, just below the Fermi level, remains constant. However, the intensity of the minority AFM peak reduces significantly for hcp-Fe/fcc-Rh in 
Fig.~\ref{fig:f3}(c) as compared to hcp-Fe/hcp-Rh in Fig.~\ref{fig:f3}(b), which describes the reduction of $\Delta E$ in the former compared to the latter.

We have identified the characteristics of the LDOS which can explain the gradual reduction of $\Delta E$, which is related to the steepness of the spin spirals curves in Fig. \ref{fig:f1}(a). The reduction of $\Delta E$ is caused by two factors: interlayer spacing and stacking order. Now, we want to examine the influence of these two factors on $\Delta E$. We find that the effect of adjusting $d_{\mathrm{Re-Re}}$ on $\Delta E$ is trivial, so we keep this interlayer distance fixed throughout. To obtain energy differences for each modification in $d_{\mathrm{Fe-Rh}}$, $d_{\mathrm{Rh-Re}}$ and in the stacking order of Fe/Rh bilayer, we introduce two intermediate ultrathin films between the three relaxed structures (Fig. \ref{fig:f4}). 

We start from the relaxed fcc/hcp geometry, which we indicated as fcc-Fe$_{2.13}$/hcp-Rh$_{2.18}$/hcp-Re [Fig. \ref{fig:f4}(a)]. The interlayer distances are expressed in \AA\ and written in subscript. We first change the overlayer stacking from fcc-Fe to hcp-Fe and obtained a system hcp-Fe$_{2.13}$/hcp-Rh$_{2.18}$/hcp-Re [Fig. \ref{fig:f4}(b)]. Then we reduce the Fe-Rh interlayer distance from 2.13 \AA\ to 2.10 \AA\ and get a new film hcp-Fe$_{2.10}$/hcp-Rh$_{2.18}$/hcp-Re [Fig.~\ref{fig:f4}(c)]. The relaxed hcp/hcp geometry is hcp-Fe$_{2.10}$/hcp-Rh$_{2.16}$/hcp-Re, which is shown in Fig. \ref{fig:f4}(d). Then, we subsequently increase the Rh-Re interlayer spacing and decrease the Fe-Re bilayer spacing by 0.04 \AA\ and get two new films$-$ hcp-Fe$_{2.10}$/hcp-Rh$_{2.20}$/hcp-Re [Fig. \ref{fig:f4}(e)] and hcp-Fe$_{2.06}$/hcp-Rh$_{2.20}$/hcp-Re [Fig. \ref{fig:f4}(f)]. Now we change the stacking of the latter film from hcp-Rh to fcc-Rh and get the relaxed fcc/hcp geometry fcc-Fe$_{2.06}$/hcp-Rh$_{2.20}$/hcp-Re [Fig. \ref{fig:f4}(g)]. We compute the total energy difference between the AFM and FM states of all the films and the results are encapsulated in Fig. \ref{fig:f5}. 

A change of Fe stacking from fcc to hcp [Fig.~\ref{fig:f5}(a) to Fig.~\ref{fig:f5}(b)] favors the AFM state and reduces $\Delta E$ by 44 meV. Then, as the Fe-Rh bilayer comes closer by 0.03 \AA\ 
[Fig.~\ref{fig:f5}(b) to Fig.~\ref{fig:f5}(c)], $\Delta E$ is further reduced by 10 meV. The next two changes [Fig.~\ref{fig:f5}(c) to Fig.~\ref{fig:f5}(d) and Fig.~\ref{fig:f5}(d) to Fig.~\ref{fig:f5}(e)] occur in the Rh-Re interlayer distance. The energy difference between the FM and AFM states almost remains constant ($\Delta E$ $\sim$ 70 meV) for changes in the Rh-Re interlayer spacing, which implies that the Fe/Rh bilayer spacing and stacking order determine the trend of $\Delta E$. Now, as we reduce the Fe-Rh bilayer spacing by 0.04 \AA\ [Fig.~\ref{fig:f5}(e) to Fig.~\ref{fig:f5}(f)], $\Delta E$ is lowered by 14 meV. A change from hcp-Rh to fcc-Rh [Fig.~\ref{fig:f5}(f) to Fig.~\ref{fig:f5}(g)] further reduces $\Delta E$ by 16 meV. As we move from relaxed hcp-Fe/hcp-Rh [Fig.~\ref{fig:f5}(d)] to relaxed fcc-Fe/hcp-Rh [Fig.~\ref{fig:f5}(g)] films, $\Delta E$ is reduced by same amount ($\sim$15 meV) due to the modification in the stacking order and interlayer spacing of Fe/Rh bilayer. Now, as we go from relaxed fcc-Fe/hcp-Rh [Fig.~\ref{fig:f5}(a)] to relaxed hcp-Fe/hcp-Rh [Fig.~\ref{fig:f5}(d)] films, the change in stacking symmetry contributes $-40$~meV and a reduction in the Fe-Rh interlayer distance costs $-10$~meV to $\Delta E$. This explains why the drop in $\Delta E$ is $\sim$ 55 meV from fcc-Fe/hcp-Rh to hcp-Fe/hcp-Rh and only $\sim$ 30 meV between hcp-Fe/hcp-Rh and hcp-Fe/fcc-Rh in Fig.~\ref{fig:f1}(a).

We have explained the trends of the three spin spiral energy dispersions of Fe/Rh/Re(0001) films based on their electronic structure. Upon including SOC, a spin spiral state becomes more favorable than the FM state in hcp-Fe/hcp-Rh, whereas the lowest energy states of fcc-Fe/hcp-Rh and hcp-Fe/fcc-Rh are ferromagnetic. Next, we investigate Rh/Fe bilayers on the Re(0001) surface.
        
\subsection{\label{sec:colmag2}Collinear magnetic states in Rh/Fe bilayer on Re(0001)}

Analogous to the discussion in Sec. \ref{sec:colmag}, we begin our investigation of Rh/Fe bilayer on Re(0001) by displaying the structural relaxed parameters based on collinear calculations [FM and $c(2\times2$) AFM states] in Table \ref{tab:table4}. We have considered all four stacking sequences: fcc-Rh/hcp-Fe, hcp-Rh/hcp-Fe, hcp-Rh/fcc-Fe and fcc-Rh/fcc-Fe. The calculations are performed using the \textsc{vasp} code.

\begin{table*}
	\centering
	\caption{Structural parameters for the four stacking sequences of the Rh/Fe bilayer on Re(0001). The interlayer distances and magnetic moments of the three outermost layer for the relaxed geometry are calculated using the \textsc{vasp} code. The energy difference between the AFM and FM states ($E_\mathrm{{AFM}}-E_\mathrm{{FM}}$) and energy of the FM/AFM states ($E_{\mathrm{relative}}^\mathrm{{FM/AFM}}$) relative to the lowest AFM state are also shown.}
	\label{tab:table4}
	\begin{ruledtabular}
		\begin{tabular}{ccccccccc} 
			Stacking & \begin{tabular}[c]{@{}c@{}}$\Delta d_{\mathrm{Rh-Fe}}$\\ (\%)\end{tabular} & \begin{tabular}[c]{@{}c@{}}$\Delta d_{\mathrm{Fe-Re}}$\\ (\%)\end{tabular} & \begin{tabular}[c]{@{}c@{}}$\Delta d_{\mathrm{Re-Re}}$\\ (\%)\end{tabular} & $\mu_{\mathrm{Rh}}$ & $\mu_{\mathrm{Fe}}$ & $\mu_{\mathrm{Re}}$ & \begin{tabular}[c]{@{}c@{}}E$_\mathrm{{AFM}}$-E$_\mathrm{{FM}}$\\ (meV/Fe)\end{tabular} & \begin{tabular}[c]{@{}c@{}}E$_{\mathrm{relative}}^\mathrm{{FM/AFM}}$\\ (meV/Fe)\end{tabular} \\
			\colrule
			fcc-Rh/hcp-Fe & $-$10.91 & $-$4.89 & $-$0.31 & 0.62 & 2.64 &  $-$0.12   & 16.4     & 10.9 (FM)\\
			hcp-Rh/hcp-Fe & $-$10.68 & $-$5.66 & $-$0.22 & 0.08 & 2.47 &  $-$0.06   & $-$61.3 & 0.00 (AFM)\\
			hcp-Rh/fcc-Fe & $-$11.67 & $-$6.60 & $-$0.99 & 0.07 & 2.44 &  $-$0.10   & $-$49.0 & 55.0 (AFM)\\
			fcc-Rh/fcc-Fe & 1.97     & $-$6.51 &    0.54 & 0.36 & 2.34 &  $-$0.08   & $-$45.1 & 567.9 (AFM)\\
		\end{tabular}
	\end{ruledtabular}
\end{table*}

The energy difference between the FM and AFM states, $\Delta E$= $E_\mathrm{{AFM}}-E_\mathrm{{FM}}$, exhibits a very interesting feature. The ground state of fcc-Rh/hcp-Fe is FM, while the other three films have a AFM ground state. All the structural parameters, magnetic moments and energy (E$_{\mathrm{relative}}^\mathrm{{FM/AFM}}$) presented in Table \ref{tab:table4} corresponding to the FM state for fcc-Rh/hcp-Fe and AFM state for the other three films. The total energies of the collinear ground states of Rh/Fe/Re(0001) films exhibit a similar trends to Fe/Rh/Re(0001) films. The AFM state of hcp-Rh/hcp-Fe has the lowest energy among all films followed by the FM state of fcc-Fe/hcp-Rh and the AFM state of hcp-Rh/fcc-Fe. The AFM state of fcc-Rh/fcc-Fe has the highest total energy among the four films. Since the relative energy of fcc-Rh/hcp-Fe and hcp-Rh/fcc-Fe films are not too high, these two along with hcp-Rh/hcp-Fe can probably be grown pseudomorphically in experiments.

Overall, the magnetic moments of Fe in Rh/Fe/Re(0001) films are small compared to Fe/Rh/Re(0001) films. This can be attributed to a larger number of nearest neighbors of Fe in Rh/Fe/Re(0001) than in Fe/Rh/Re(0001). The magnetic moments of Fe is 2.6 $\mu_{\mathrm{B}}$ in fcc-Rh/hcp-Fe/Re(0001), 2.3 $\mu_{\mathrm{B}}$ in fcc-Rh/fcc-Fe/Re(0001) and 2.4 $\mu_{\mathrm{B}}$ in hcp-Rh/fcc-Fe/Re(0001) and fcc-Rh/hcp-Fe/Re(0001). Note that the magnetic moments of Fe in Rh/Fe/Ir(111) ultrathin films are 2.4 $\mu_{\mathrm{B}}$ \cite{rhfeir111}. The induced moments on Rh in Rh/Fe/Re(0001) films varies from 0.1 $\mu_{\mathrm{B}}$ to 0.6 $\mu_{\mathrm{B}}$ , while it is $~$0.3 $\mu_{\mathrm{B}}$ in Rh/Fe/Ir(111). The Rh layer of the Fe/Rh/Re(0001) systems possesses an induced moment of $\sim$0.1 $\mu_{\mathrm{B}}$. The small magnetic moments of Fe in Rh/Fe/Re(0001) films as compared to Fe/Rh/Re(0001) films are caused by the large Rh-Fe relaxation. 

For the first three films in Table \ref{tab:table4}, fcc-Rh/hcp-Fe, hcp-Rh/hcp-Fe and hcp-Rh/fcc-Fe, we see nearly $\sim$11$\%$ inward relaxation of Rh overlayer. The Fe-Re interlayer relaxation ($\Delta d_{\mathrm{Fe-Re}}$) is inward and the values vary from 4.9$\%$ to 6.6$\%$. The first Re-Re layer distance reduces after relaxation for the first three films. The value is nearly the same for fcc-Rh/hcp-Fe and hcp-Rh/hcp-Fe ($-$0.3 and $-$0.2) and slightly larger for hcp-Rh/fcc-Fe ($-$1.0). Similar to Rh/Fe/Re(0001) films, nearly 11$\%$ inward relaxation of Rh overlayer in hcp-Rh/Fe/Ir(111) and fcc-Rh/Fe/Ir(111) has been reported \cite{rhfeir111}. Analogous to our films, the Fe-Ir interlayer and Ir-Ir interlayer relaxation of Rh/Fe/Ir(111) is inward and the values are $\sim$6$\%$ and $\sim$0.3$\%$, respectively. The same structural relaxed parameters of Rh/Fe/Re(0001) and Rh/Fe/Ir(111) could prompt one to naively conclude that the Rh/Fe bilayer mainly controls the properties and the substrate does not have a notable effect. However, we see below that the substrate has a significant effect on the ground state and we discuss it in the next section.   

The relaxed parameters of fcc-Rh/fcc-Fe/Re(0001) are different compared to the other three films. The relaxation of the Rh-Fe interlayer is outward and the value is strikingly small. The magnitude and sign of the Fe-Re layer relaxation show similarities with the other films, however, the Re-Re interlayer relaxation is opposite. We exclude this film since it has an extremely large relative energy. We also exclude hcp-Rh/fcc-Fe/Re(0001) since its behavior would be similar to that of hcp-Rh/hcp-Fe/Re(0001), which we have already seen for their counterparts in Fe/Rh/Re(0001).

\begin{figure}[!h]
	\includegraphics[scale=1.0]{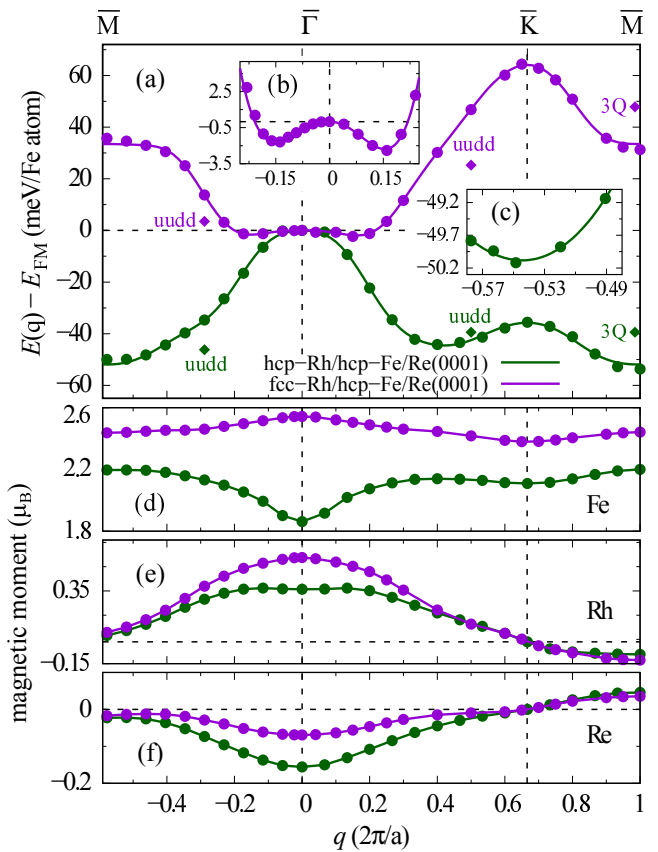}
	\centering
	\caption{(a) Energy dispersions of flat spin spirals along two high symmetry directions ($\overline{\Gamma \mathrm{KM}}$ and $\overline{\Gamma \mathrm{M}}$) for fcc-Rh/hcp-Fe/Re(0001) (violet) and hcp-Rh/hcp-Fe/Re(0001) (green). The filled circles represent DFT data and the solid lines are the fit to the Heisenberg model. The filled diamonds represent the 3$Q$ and $uudd$ states. Inset (b) shows energy dispersion of fcc-Rh/hcp-Fe/Re(0001) around $\overline{\Gamma}$ and (c) displays energy dispersion of hcp-Rh/hcp-Fe/Re(0001) around the $\overline{\mathrm{M}}$ point along $\overline{\Gamma \mathrm{M}}$ direction. Magnetic moments of Fe, Rh and Re (third layer) are shown in panel (d), (e) and (f), respectively.} 
	\label{fig:f6}
\end{figure}

\subsection{\label{sec:ncolmag2}Noncollinear magnetism in Rh/Fe bilayer on Re(0001)}

\begin{table*}[!thbp]
	\centering
	\caption{Effective exchange constant ($J_{\mathrm{eff}}$), Exchange parameters ($J_{ij}$), biquadratic exchange constant ($B_{1}$), 3-site four spin exchange constant ($Y_{1}$) and 4-site four spin exchange constant ($K_{1}$) for two stackings of Rh/Fe bilayer on Re(0001). $\Delta E$ is the energy difference between the multi-$Q$ and spin spiral states  as defined by Eqs.~(9-11).
  All energies are given in meV.}
	\label{tab:table5}
	\begin{ruledtabular}
		\begin{tabular}{ccccccccccccccc}
			Stacking & $J_{\mathrm{eff}}$ & $J_{1}$ & $J_{2}$ & $J_{3}$ & $J_{4}$ & $J_{5}$ & $J_{6}$ & $J_{7}$ & $B_{1}$ & $Y_{1}$ & $K_{1}$ & $\Delta E_{\overline{\mathrm{M}}}^{3Q}$ & $\Delta E_{3\overline{\mathrm{K}}/4}^{uudd}$ & $\Delta E_{\overline{\mathrm{M}}/2}^{uudd}$\\
			\colrule
			fcc-Rh/hcp-Fe & $-$2.5 & 6.85 & $-$1.68 & 0.67 & $-$0.27 & $-$1.02 & $-$0.02 & 0.08 & 2.83 & $-$1.41 & $-$0.57 & 16.52 & $-$21.49 & $-$10.22\\
			hcp-Rh/hcp-Fe & $-$0.3 & $-$4.13 & $-$2.40 & $-$0.13 & $-$0.04 & $-$0.44 & 0.17 & 0.10 & 2.78 & 1.82 & 0.86 & 14.26 & 3.05 & $-$11.54\\
		\end{tabular}
	\end{ruledtabular}
\end{table*}

To check whether the collinear magnetic state is the true ground state or a spin spiral state can be more favorable, we have calculated the energy dispersion of flat homogeneous spin spirals along the $\overline{\Gamma \mathrm{KM}}$ and $\overline{\Gamma \mathrm{M}}$ direction of the 2DBZ for hcp-Rh/hcp-Fe and fcc-Rh/hcp-Fe using \textsc{fleur}. The energies of the multi-$Q$ states are also calculated for consistency. The results are displayed in Fig.~\ref{fig:f6}.

First notice that the behavior of the two energy dispersion curves are qualitatively different and it is surprising that a change of the overlayer stacking can cause such a drastic effect on the magnetic interactions (Fig.~\ref{fig:f6}(a)). We see that, for these two films, a spin spiral configuration has a lower energy than the collinear state. The spin spiral minimum is observed along $\overline{\Gamma \mathrm{M}}$ direction for hcp-Rh/hcp-Fe. The energy minimum is located at $q$= 0.03$(\frac{2\pi}{a}$) with respect to the $\overline{\mathrm{M}}$ point, which corresponds to a wavelength of 9.2 nm, and it is 0.3 meV/Fe lower than the 
{$\overline{\mathrm{M}}$} point. For fcc-Rh/hcp-Fe, a spin spiral minimum in both high symmetry directions is observed, however, the energy is lowest in the $\overline{\Gamma \mathrm{KM}}$ direction. The energy minimum of $E_{\mathrm{min}}$= $-$2.4 meV/Fe occurs at $q$= 0.16$(\frac{2\pi}{a}$), which corresponds to a wavelength of 1.7 nm. 

The $uudd$ state along $\overline{\Gamma \mathrm{M}}$ is only $\sim$3.5 meV higher than the FM state in fcc-Rh/hcp-Fe and it is higher than the $\overline{\mathrm{M}}$ point by the same small amount in hcp-Rh/hcp-Fe. The $uudd$ state along $\overline{\Gamma \mathrm{KM}}$ and the 3$Q$ state resides at a quite high energy compared to the FM state. In contrast to our observation in Rh/Fe/Re(0001), the two dispersion curves corresponding to the hcp and fcc stacking of Rh in Rh/Fe/Ir(111) behave qualitatively in a similar way and exhibit a pronounced spin spiral minimum \cite{rhfeir111}. However, the $uudd$ state of hcp-Rh/Fe/Ir(111) along $\overline{\Gamma \mathrm{M}}$ is energetically lower than the spin spiral state and becomes the ground state of this film. The differences in the behavior of the spin spiral dispersion curves and in the energetics of the multi-$Q$ states between Rh/Fe/Re(0001) and Rh/Fe/Ir(111) are caused by substrate. The reason for this large effect will be discussed in the next section. 

The magnetic moments of individual layers for fcc-Rh/hcp-Rh/Re(0001) and hcp-Rh/hcp-Fe/Re(0001) are shown in Fig. \ref{fig:f6}(b-d). The magnetic moments of Fe in hcp-Rh/hcp-Fe varies by 0.4 $\mu_{\mathrm{B}}$ over the whole range of $\textbf{q}$. The variation is large around the FM state and the moments become nearly constant in the vicinity of both $\overline{\mathrm{M}}$ points, which are our areas of interest. The Fe magnetic moments at the $\overline{\mathrm{M}}$ point is 2.2 $\mu_{\mathrm{B}}$ for hcp-Rh/hcp-Fe. The moments of Fe are fairly constant throughout the range of $\textbf{q}$ for fcc-Rh/hcp-Fe. The induced moments of Rh and Re also vary with $\textbf{q}$ around the FM state, however, it becomes almost constant close to the $\overline{\mathrm{M}}$ points. 

The exchange parameters are computed by projecting the spin spiral energies of Fig. \ref{fig:f6}(a) onto the Heisenberg hamiltonian [Eq. (\ref{eq1})] and the calculated exchange constants are tabulated in Table \ref{tab:table5}. The effective exchange constant is obtained from a fit in the vicinity of the $\overline{\Gamma}$ point for fcc-Rh/hcp-Fe and in the vicinity of the $\overline{\mathrm{M}}$ point for hcp-Rh/hcp-Fe. The negative sign of the effective exchange constants indicates that the magnetic ground state of fcc-Rh/hcp-Fe and hcp-Rh/hcp-Fe has a spin spiral configuration. The exchange parameters of both the films demonstrate a frustrated nature which is consistent with the observation of a spin spiral ground state in the dispersion curve [Fig. \ref{fig:f6}(a)].    

The energy of the multi-$Q$ states with respect to the corresponding spin spiral state and the higher-order exchange parameters for fcc-Rh/hcp-Fe and hcp-Rh/hcp-Fe are listed in Table \ref{tab:table5}. Both films have nearly the same biquadratic constants ($B_{1}$$\sim$ 2.7 meV), while the 3-site four spin ($Y_{1}$) and 4-site four spin exchange ($K_{1}$) constants have different values depending on the stacking of the Rh/Fe bilayer. The energy separation of the 3$Q$ state from $\overline{\mathrm{M}}$ point along $\overline{\Gamma \mathrm{KM}}$, $\Delta E_{\overline{\mathrm{M}}}^{3Q}$, and the energy seperation of the $uudd$ state from $\textbf{q}$=$\overline{\mathrm{M}}/2$ spiral state, $\Delta E_{\overline{\mathrm{M}}/2}^{uudd}$, remain almost unchanged for both films. This explains why the biquadratic constants do not depend on the stacking order of Rh/Fe bilayers. The sign of $\Delta E_{3\overline{\mathrm{K}}/4}^{uudd}$ changes based on the stacking order of Rh/Fe bilayers, which is reflected in the value of the 3-site and 4-site four spin exchange constants. Now, it is straightforward to understand the magnitude of these two HO exchange constants using Eqs.~(\ref{eq7}-\ref{eq8}).             

We have seen that SOC promotes a spin spiral state by making it lower than the FM state in hcp-Fe/hcp-Rh/Re(0001). Here, due to exchange frustration, a spin spiral state has already becomes the ground state for fcc-Rh/hcp-Fe and hcp-Rh/hcp-Fe without SOC. Now we see the impact of SOC on spin spiral states.

The MAE of fcc-Rh/hcp-Fe and hcp-Rh/hcp-Fe are listed in Table \ref{tab:table6}. The MAE favors an in-plane magnetization direction for both the films and the magnitude is relatively higher than their counterpart in Fe/Rh/Re(0001). The MAE constant of fcc-Rh/hcp-Fe accounts to $\sim$ 2 meV and it is 1.6 meV in hcp-Rh/hcp-Fe. A similar trend of MAE with the stacking of Rh layer has been observed in Rh/Fe/Ir(111), however, the easy axis is out-of-plane \cite{rhfeir111}.

\begin{table} [!htbp]
	\centering
	\caption{Effective Dzyaloshinskii-Moriya interaction constant ($D_{\mathrm{eff}}$), magnetocrystalline anisotropy ($K_{\mathrm{MAE}}$) for two stackings of Rh/Fe bilayer on Re(0001). The negative value of $K_{\mathrm{MAE}}$ indicates an in-plane easy magnetization axis. All energies are given in meV.}
	\label{tab:table6}
	\begin{ruledtabular}
		\begin{tabular}{ccc}
			Stacking & $D_{\mathrm{eff}}$ & $K_{\mathrm{MAE}}$  \\
			\colrule
			fcc-Rh/hcp-Fe & 1.19 & $-$2.01  \\
			hcp-Rh/hcp-Fe & 0.58 & $-$1.61 \\
		\end{tabular}
	\end{ruledtabular}
\end{table}

\begin{figure}[h!]
	\includegraphics[scale=1.0]{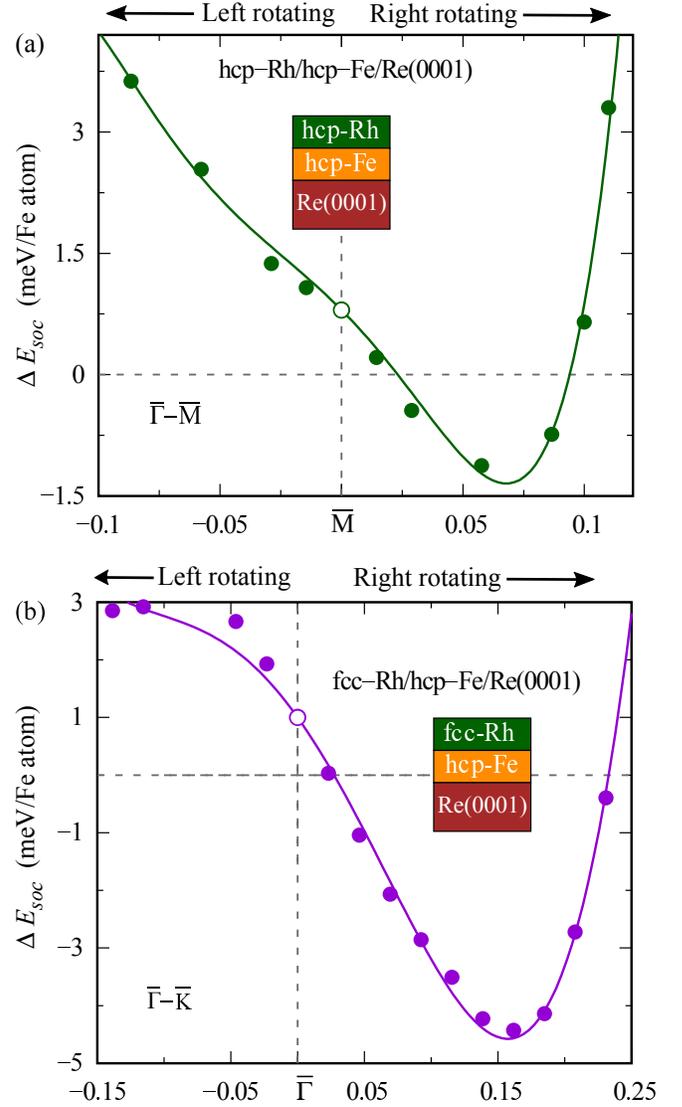} 
	\centering
	\caption{Energy dispersions of flat cycloidal spin spirals including DMI and MAE (a) along $\overline{\Gamma \mathrm{M}}$ for hcp-Rh/hcp-Fe/Re(0001) (green) and (b) along $\overline{\Gamma \mathrm{K}}$ for fcc-Rh/hcp-Fe/Re(0001) (violet). The filled circles represent DFT data and the solid lines are fits to the Heisenberg [Eq. (\ref{eq1})] plus the DMI [Eq. (\ref{eq2}] models. MAE shifts the energy by $K_{\mathrm{MAE}}$/2 with respect to the $\overline{\Gamma}$ or $\overline{\mathrm{M}}$} points. 
	\label{fig:f7}
\end{figure}

The spin spiral energy dispersion of hcp-Rh/hcp-Fe and fcc-Rh/hcp-Fe including the contributions from the DMI and MAE is displayed in Fig. \ref{fig:f7}. The symmetry direction is chosen to exhibit the maximum effect of SOC. The DMI constant is extracted by fitting the total SOC energy in the vicinity of the $\overline{\Gamma}$ point for fcc-Rh/hcp-Fe and close to the $\overline{\mathrm{M}}$ point for hcp-Rh/hcp-Fe. We have seen that the major DMI contribution in Fe/Rh/Re(0001) comes from the 3$d$-4$d$ (Fe-Rh) interface. However, in Rh/Fe/Re(0001), the 3$d$-5$d$ (Fe-Re) interface provides the largest contribution to SOC followed by the 4$d$-3$d$ (Rh-Fe) interface. In case of fcc-Rh/hcp-Fe, both interfaces (Fe-Re and Rh-Fe) prefer right-rotating (clockwise) spin spirals and therefore, enhance the strength of DMI ($D_{\mathrm{eff}}$= 1.19 meV), evident from a comparison with the DMI constants of fcc-Fe/hcp-Rh/Re(0001). In hcp-Rh/hcp-Fe, the two interfaces favor spin spirals with opposite rotational sense and thus reduce the DMI strength ($D_{\mathrm{eff}}$= 0.58 meV). The DMI makes the spin spiral minimum more pronounced for both films and moves the energy minimum of hcp-Rh/hcp-Fe to $q$= 0.07$(\frac{2\pi}{a}$) ($\lambda$$\approx$ 4 nm), whereas the energy minimum of fcc-Rh/hcp-Fe remains at $q$= 0.16$(\frac{2\pi}{a}$) ($\lambda$$\approx$ 1.7 nm). The energy minimum is 1.5 meV and 5 meV below the FM state for hcp-Rh/hcp-Fe and fcc-Rh/fcc-Fe, respectively. The depth of energy minima indicates that isolated skyrmions can only be stabilized at extremely large magnetic fields, which could be obtained by an exchange bias effect of an adjacent magnetic layer \cite{Nandy2016}.

\begin{figure}[!htbp]
	\includegraphics[scale=1.0]{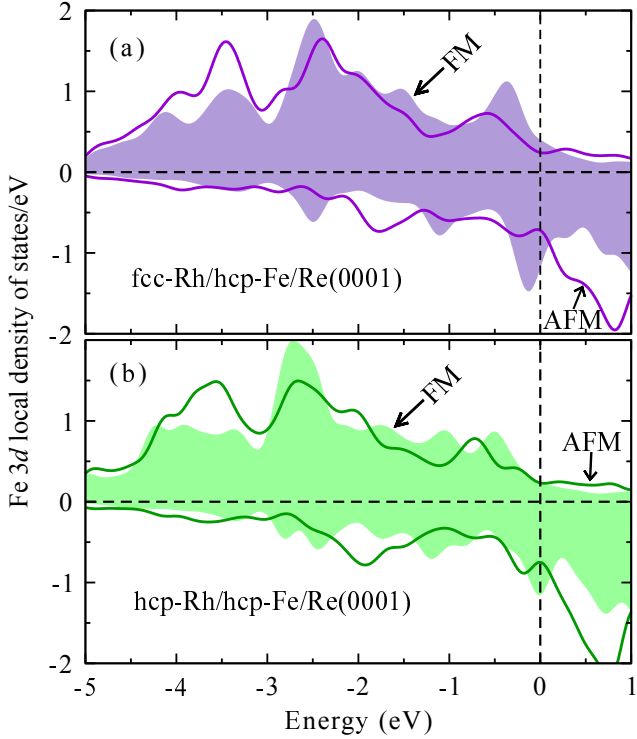}
	\centering
	\caption{Spin-polarized Fe 3$d$ local density of states (LDOS) for (a) fcc-Rh/hcp-Fe/Re(0001) (violet) and (b) hcp-Rh/hcp-Fe/Re(0001) (green). FM (filled) and AFM (line) LDOS are shown in each panels.} 
	\label{fig:f8}
\end{figure}

\subsection{\label{sec:estr2} Trends of the spin spiral curves from electronic structures} 

\begin{figure}[!htbp]
	\includegraphics[scale=1.0]{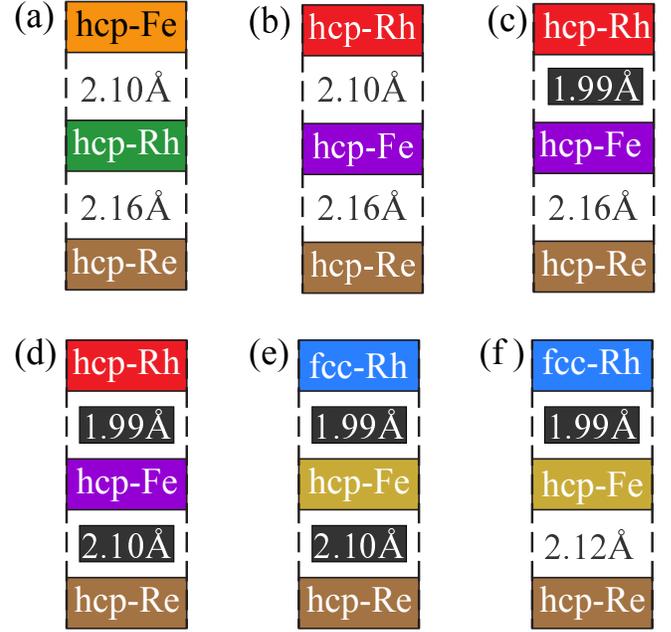}
	\centering
	\caption{Rh/Fe/Re(0001) films with various interlayer distances and stacking sequences.  First three layers are shown for convenience. Films (a), (d) and (f) represents a relaxed geometry, i.e., i.e., their interlayer distances along with the order of Rh/Fe bilayer are consistent with Table \ref{tab:table4}. We introduce two intermediate films between (a)-(d) and one film between (d)-(f) to study the change of $\Delta E$=$E_{\mathrm{AFM}}-E_{\mathrm{FM}}$. Film (a) corresponds to Fig. \ref{fig:f4}(d). Each modification in the interlayer distance and stacking order is denoted by a change of color.} 
	\label{fig:f9}
\end{figure}

To achieve a qualitative understanding regarding the drastic change in relative energies of the collinear states in fcc-Rh/hcp-Fe and hcp-Rh/hcp-Fe, as reflected in the sign of $\Delta E$=$E_{\mathrm{AFM}}-E_{\mathrm{FM}}$, we compute the LDOS of Fe atom in the FM and AFM states for both the films [Fig.~\ref{fig:f8}(a-b)]. Each panel displays the Fe 3$d$ LDOS in the FM and AFM state. In the top panel, we show the LDOS for fcc-Rh/hcp-Fe and in the bottom panel it is allotted to hcp-Rh/hcp-Fe.

The minority FM peak of hcp-Rh/hcp-Fe, which resides at the Fermi level, moves to lower energies in fcc-Rh/hcp-Fe. A peak at the Fermi level signifies instability and its movement to the lower energy makes the FM state more stable in fcc-Rh/hcp-Fe [Fig.~\ref{fig:f8}]. The majority AFM LDOS between $-2$ eV and $-4.5$ eV moves to the higher energy relative to the FM LDOS in fcc-Rh/hcp-Fe as compared to hcp-Rh/hcp-Fe. We observe a significant increment of the majority AFM LDOS and a large decrement of the majority FM LDOS in fcc-Rh/hcp-Fe around $-2.5$ eV compared to the other film. These changes in the
LDOS provide a positive energy contribution to $\Delta E$ and lower the energy of the FM state. The intensity of the majority FM peak of fcc-Rh/hcp-Fe just below the Fermi level increases compared to hcp-Rh/hcp-Fe, which acts opposite to the other features and tries to increase the energy seperation between the collinear states of hcp-Rh/hcp-Fe. The magnitude of the positive energy contribution is larger than the negative one which makes the FM state lower in energy than the AFM state in fcc-Rh/hcp-Fe.     

\begin{figure}[!htbp]
	\includegraphics[scale=1.0]{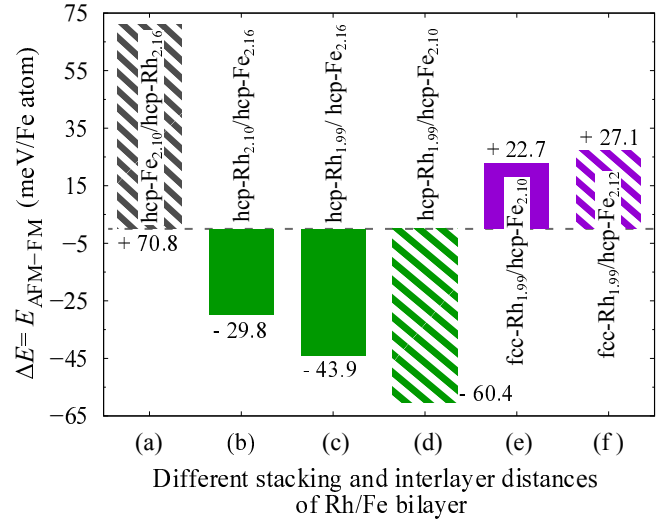}
	\centering
	\caption{Energy difference between the AFM and FM states ($\Delta E$) for films of Fig. \ref{fig:f9}. Films with relaxed geometry [(a), (d) and (f)] are shown using shaded color and intermediate films [(b), (c) and (e))] are shown using solid color.} 
	\label{fig:f10}
\end{figure}

To obtain a quantitative understanding of the relative energy difference $\Delta E $=E$_{\mathrm{AFM}}-$E$_{\mathrm{FM}}$ in fcc-Rh/hcp-Fe and hcp-Rh/hcp-Fe, we systematically modify the interlayer distances ($d_{\mathrm{Rh-Fe}}$ and $d_{\mathrm{Fe-Re}}$) and stacking sequence of Rh/Fe bilayer and study the energy difference between the FM and AFM states. 

To identify the effect of layer swapping, we begin from the relaxed geometry of hcp-Fe/hcp-Rh/Re(0001). We introduce two intermediate ultrathin films between the relaxed geometry of hcp-Fe/hcp-Rh/Re(0001) and hcp-Rh/hcp-Fe/Re(0001) and one intermediate ultrathin film between the relaxed geometry of hcp-Rh/hcp-Fe/Re(0001) and fcc-Rh/hcp-Fe/Re(0001).  

The relaxed geometry of hcp-Fe/hcp-Rh/Re(0001) is represented as hcp-Fe$_{2.10}$/hcp-Rh$_{2.16}$/hcp-Re [Fig. \ref{fig:f9}(a)]. First we swap the Fe-Rh layer of hcp-Fe/hcp-Rh/Re(0001) and get hcp-Rh$_{2.10}$/hcp-Fe$_{2.16}$/hcp-Re [Fig. \ref{fig:f9}(b)]. Then we reduce the Rh-Fe interlayer distance from 2.10 \AA\ to 1.99 \AA\ and obtain hcp-Rh$_{1.99}$/hcp-Fe$_{2.16}$/hcp-Re [Fig. \ref{fig:f9}(c)]. The relaxed hcp-Rh/hcp-Fe geometry is hcp-Rh$_{1.99}$/hcp-Fe$_{2.10}$/hcp-Re, which is shown in [Fig. \ref{fig:f9}(b)]. Now we change the stacking of the overlayer from hcp to fcc and get a new film fcc-Rh$_{1.99}$/hcp-Fe$_{2.10}$/hcp-Re [Fig. \ref{fig:f9}(e)]. Then we increase the Fe-Re interlayer spacing from 2.10 \AA\ to 2.12 \AA\ and get the relaxed geometry of fcc-Rh/hcp-Fe as fcc-Rh$_{1.99}$/hcp-Fe$_{2.12}$/hcp-Re in Fig. \ref{fig:f9}(f).

First notice that a swapping of the topmost two layers [Fig.~\ref{fig:f10}(a) to Fig.~\ref{fig:f10}(b)] changes the relative stability of the two collinear states. The AFM state which is energetically higher than the FM state in Fig.~\ref{fig:f10}(a) becomes energetically lower in Fig.~\ref{fig:f10}(b). Based on DFT calculations, it has been shown~\cite{hardrat} that the ground state prefers a FM configuration for an Fe monolayer on Rh(111) and an AFM configuration for an Fe monolayer on Re(0001), which implies a competition of two opposite mechanism in the films of Fig.~\ref{fig:f10}(b). It is also observed in Ref.~\cite{hardrat} that the strength of the AFM interaction in Fe/Re(0001) is four times stronger than the FM interaction in Fe/Rh(111), which clarifies the reason for observing a AFM ground state in Fig.~\ref{fig:f10}(b). The total energy changes by a sizable amount ($\sim$100 meV) due to the swapping [Fig.~\ref{fig:f10}(a) to Fig.~\ref{fig:f10}(b)]. 

As we reduce the Rh-Fe interlayer distance by 0.1 \AA\ [Fig.~\ref{fig:f10}(b) to Fig.~\ref{fig:f10}(c)], $\Delta E$ becomes more negative by 14 meV. A reduction of the Fe-Re distance by 0.04 \AA\ [Fig. \ref{fig:f10}(c) to Fig. \ref{fig:f10}(d)] corresponds to a energy change of $-$16.5 meV, which implies that the Fe-Re hybridization is quite strong. Now, the AFM state nearly gains 83 meV energy, as we adjust the stacking of Rh overlayer from hcp to fcc [Fig.~\ref{fig:f10}(d) to Fig.~\ref{fig:f10}(e)] and becomes $\sim$ 23 meV higher than the FM state. Considering the energy gain of nearly 44 meV from hcp-Fe/Rh/Re(0001) to fcc-Fe/Rh/Re(0001), we argue that the strong hybridization between the Fe and Re layers plays an indirect role for such a large energy increment. From Fig.~\ref{fig:f10}(e) to 
Fig.~\ref{fig:f10}(f), an increment of Fe-Rh interlayer spacing by 0.02 \AA\ increase the relative energy of the AFM state by 5 meV. 

\section{\label{sec:conc} Conclusion}
We have investigated the thermodynamic stability of different stacking orders of Fe/Rh and Rh/Fe bilayers on Re(0001) using DFT. Our study incorporates all four possible stacking (hcp and fcc) combinations of Fe/Rh and Rh/Fe bilayers. We calculate the exchange parameters and an effective DMI constant by mapping the spin spiral and SOC energies onto the respective spin Hamiltonian, whereas the MAE and higher-order exchange constants are computed directly from DFT. We observe that the  magnetic interactions depend strongly on the stacking order of the topmost two layers. We find that the hcp/hcp staking sequence (hcp-Fe/hcp-Rh and hcp-Rh/hcp-Fe) has the most stable pseudomorphic structure for both cases. An energy comparison of all types of stackings indicates that the fcc/hcp, hcp/fcc and hcp/hcp type structures can be grown pseudomorphically in the experiment. On the other hand, the fcc/fcc type structure can barely be realized in the experiment. Note that a pseudomorphic growth of an
Fe monolayer on the Re(0001) surface has already been observed in experiments \cite{fere0001e}. 
        
The spin spiral dispersion of Fe/Rh/Re(0001) films show that the exchange interaction favors a FM ground state. A strong hybridization of Rh with Fe as compared to a weak one with Re can explain the trend in the slope of the dispersion curves. Upon including SOC, a right-rotating spin spiral with a period of nearly 11~nm becomes favorable than the FM state in hcp-Fe/hcp-Rh/Re(0001), while the FM state remains the ground state for fcc-Fe/hcp-Rh/Re(0001) and hcp-Fe/fcc-Rh/Re(0001). We do not observe any significant effect of the higher-order exchange interactions on the ground state in contrast
to the observations for Rh/Fe bilayers on Ir(111).
The energy dispersion profile of hcp-Fe/hcp-Rh/Re(0001) indicates that isolated skyrmions can be stabilized in this system. Indeed in atomistic spin dynamics simulations for this system, we observe isolated skyrmions at an external magnetic field~\cite{mine}. In that work, we also demonstrate that the stability of skyrmions is strongly affected by higher-order interactions.

The spin spiral dispersion of fcc-Rh/hcp-Fe/Re(0001) and hcp-Rh/hcp-Fe/Re(0001) exhibit an exchange-driven spin spiral ground state. The spin spiral minimum of the former is located close to the FM state and the minimum of the later is located close to the AFM state. The robust hybridization of Fe with Rh and Re layers is responsible for drastic variation in the dispersion profile against the change of overlayer symmetry. The SOC contribution further lowers the energy of the spin spirals, which implies that isolated skyrmions can only be stabilized at extremely large magnetic fields which can
be obtained by exploting an exchange bias effect with an adjacent magnetic layer.

The variety of complex magnetic structures predicted based on our first-principles calculations suggests that experiments on bilayers of Fe and Rh on Re(0001) are promising. In particular, the possibility
to observe magnetic skyrmions in an ultrathin film with an easy in-plane magnetization is intriguing.

\bibliography{reference}
\end{document}